\documentstyle[12pt]{article}
\newcommand{\nn}{\nonumber\\}\newcommand{\p}[1]{(\ref{#1})}
\begin{document}
\thispagestyle{empty}
\renewcommand{\thefootnote}{\dagger}

\hbox to \hsize{\hfill G\"oteborg-ITP-94-10}
\hbox to \hsize{\hfill hep-th/9403181}
\vskip1.5truecm
\centerline
{\bf TOWARDS A COMPLETE TWISTORIZATION}
\centerline{\bf OF THE HETEROTIC STRING}

\vskip 1.5truecm

\centerline
{\bf Igor A. Bandos}

\bigskip
\centerline{\sl National Research Centre}
\centerline{\sl Kharkov Institute of Physics and Technology}
\centerline{\sl Kharkov, 310108 Ukraine}

\bigskip
\centerline
{\bf
Martin Cederwall\footnote{email address: tfemc@fy.chalmers.se}
\hskip24pt\renewcommand{\thefootnote}{\ddagger}
Dmitrij P. Sorokin
\footnote{Permanent address: Kharkov Institute of Physics and Technology,
	Kharkov, 310108, the Ukraine,}\renewcommand{\thefootnote}{}
\footnote{e-mail address: dsorokin@kfti.kharkov.ua}}

\bigskip
\centerline{\sl Chalmers University of Technology and University of Gothenburg}
\centerline{\sl Institute for Theoretical Physics}
\centerline{\sl S-412 96 Gothenburg, Sweden}

\medskip
\centerline{\bf and}

\medskip
\centerline
{\bf Dmitrij V. Volkov}

\bigskip
\centerline{\sl National Research Centre}
\centerline{\sl Kharkov Institute of Physics and Technology}
\centerline{\sl Kharkov, 310108 Ukraine}


\vskip 2truecm

\noindent
{\bf Abstract.} In $D=3,4,6$ and 10 space--time dimensions considered is a
string model invariant under transformations of $N=1$ space--time supersymmetry
and $n=D-2$ local worldsheet supersymmetry with the both Virasoro constraints
solved in the twistor form. The twistor solution survives in a modified form
even in the presence of the heterotic fermions.

\vskip1truecm

\setcounter{page}0
\renewcommand{\thefootnote}{\arabic{footnote}}
\setcounter{footnote}0
\newpage
\section{Introduction}
Recently progress has been made in understanding the geometrical origin of the
fermionic $\kappa$--symmetry \cite{ks} of $D=3,4,6$ and 10 Brink--Schwarz
superparticles \cite{bs}, and Green--Schwarz superstrings \cite{gsw}. This has
been achieved through the development of a formulation of these theories
possessing space--time supersymmetry and local $n=D-2$ worldsheet supersymmetry
the latter turning to the $\kappa$--symmetry on the mass shell \cite{stv}. Such
a doubly supersymmetric formulation provides a natural ground for the twistor
transform \cite{lli} in describing superparticle and superstring dynamics
\cite{stw} since twistor components (being commuting spinors) arise as
superpartners of the Grassmann coordinates of target superspace. The motion of
the superparticle and superstring determines the embedding of the worldsheet
supersurface into the target superspace (and vice versa) in such a way that for
dynamical characteristics of the objects the Cartan--Penrose representation  of
the light--like vector as a bilinear combination of spinors arises.
A hope has been that this approach may contain the advantages of both the
Green--Schwarz and the Neveu--Schwarz--Ramond formulation of the superstrings
and at the same time will be free of their problems.
(On closely related Lorentz--harmonic approaches see \cite{harm}, \cite{bz}).

Up to now the twistor--like formulation has been developed for describing
$D=3,4,6$ and 10 $N=1$ superparticles \cite{stv}, \cite{spa}, \cite{n8p}, $N=1$
null superstrings \cite{bstv} and $N=1$ heterotic strings
\cite{sstr}--\cite{dg} though the latter encountered the problem of
incorporating  chiral fermions \cite{st}. At the same time a problem arose
with generalizing these results to construct $n=D-2$ worldsheet superfield
versions of the $N=2$ superstrings \cite{vz}--\cite{gs2}. The difficulty is to
not allow the auxiliary fields in the superfield action of the string, such as
Lagrange multipliers, to propagate. For $D=3$, type IIA superstring it was
solved in \cite{gs2}, while for getting an appropriate D=10, type IIA
superstring action the authors of ref. \cite{ptm} proposed to perform a
dimensional reduction of a $D=11$,
$N=1$ twistor--like supermembrane theory. This problem of the manifest doubly
supersymmetric description of $N=2$ dynamics is apparently connected with the
problem of solving for the both Virasoro constraints in the twistor form.
Component actions of this kind for describing $N=1$ as well as $N=2$
superstrings have been studied in \cite{ps}, and relevant superfield equations
of
motion of $N=2$, $D=4$ superstring  were discussed in \cite{vz}. At the same
time all known superfield versions of $D=3,4,6$ and 10 $N=1$ superstrings deal
with the twistorization of only one Virasoro constraint corresponding to the
supersymmetric (for example, the right--moving) sector of the heterotic string,
while the Virasoro constraint corresponding to the left--moving sector is taken
into account in the ``old--fashion'' manner.

In the present paper we propose a  doubly twistorized supersymmetric version of
an $N=1$ superstring in $D=3,4,6$ and 10 space--time. We will also show that
inspite of the fact that the left--moving (non--supersymmetric) sector
of the heterotic string is a relic of a  bosonic string in $D=26$ \cite{gsw},
where the conventional twistor relation does not take place, it is nevertheless
possible to keep the twistor solution to the corresponding Virasoro constraint
in the presence of the chiral fermions.

The aim of the paper is twofold. From the one hand side it seems of interest to
put the both Virasoro constraints of the heterotic string onto an equal
``twistor'' footing, and from the other hand  the construction of this version
testifies, though indirectly, to the existence of the $n=D-2$ worldsheet
superfield formulation of type IIA \cite{ptm}, \cite{gs2} and IIB superstrings,
since the former may be thought of as a truncated theory of the latter.

\section{Geometry of $(0,D-2)$ worldsheet superspace}

To construct a superfield action one has to specify the geometrical properties
of the superspace where the superfields of the theory are determined.

In the case under consideration we deal with a worldsheet superspace
parametrized by two bosonic and $n=D-2~ (D=3,4,6,10)$ fermionic coordinates
$z^{\cal M}=(\xi^\mu, \eta^A)$, where $\mu=0,1$; and $A=1,2,...,n$ are to be
identified with indices of an n-dimensional representation of an internal
$SO(n)$ group. For physical reasons the objects such as supervielbeins
$E^{\cal A} =dz^{\cal M}E^{\cal A}_{\cal M}$, $SO(1,1)\times SO(n)$
superconnection $\Omega= dz^{\cal M}\Omega_{\cal M}$, torsion $T^{\cal
A}=E^{\cal B}\wedge E^{\cal C}T^{\cal A}_{{\cal B}{\cal C}}={\cal D}E^{\cal A}$
and curvature $R_{\cal B}^{~\cal A}=
E^{\cal D}\wedge E^{\cal C}R_{{\cal D}{\cal C}{\cal B}}^{~~~\cal A}=
d\Omega_{\cal B}^{~\cal A}+\Omega_{\cal B}^{~\cal C}\wedge\Omega_{\cal
C}^{~\cal A}$ determining supergeometry should be constrained.\footnote{${\cal
D}=d+\Omega= E^{\cal A}(\nabla_{\cal A}+\Omega_{\cal A})= E^{\cal A}{\cal
D}_{\cal A}$ denotes a covariant exterior derivative, and ${\cal A},{\cal
B},{\cal C}$ stand for
$(++,--,-{A})$, where the indices of the $SO(n)$ structure subgroup are
distinguished from the $\eta^{A}$--coordinate indices by the sign $(-)$
indicating the conformal weight ${1\over 2}$ of corresponding right--moving
spinors.}
This is achieved by imposing the following conditions \cite{cons}, \cite{t}:
\begin{equation}\label{1}
T^{++}=0,
\end{equation}
 \begin{equation}\label{2}
T^{--}=iE^{-A}\wedge E^{-A},
\end{equation}
 \begin{equation}\label{3}
T^{-A}=E^{--}\wedge E^{++}T^{-A}_{--++},
\end{equation}
or in terms of the covariant derivatives:
\begin{equation}\label{4}
[{\cal D}_{++},{\cal D}_{-A}]=R_{++-A},
\end{equation}
 \begin{equation}\label{5}
[{\cal D}_{--},{\cal D}_{-A}]=0,
\end{equation}
 \begin{equation}\label{6}
\{{\cal D}_{-A},{\cal D}_{-B}\}=-2i\delta_{AB}{\cal D}_{--},
\end{equation}
 \begin{equation}\label{7}
[{\cal D}_{--},{\cal D}_{++}]=T^{-A}_{--++}{\cal D}_{-A}+R_{--++}.
\end{equation}

In \p{4}--\p{7} we skipped the $SO(n)$ curvature components since when
constructing the twistor superstring action we shall deal with the superfields
being singlets of $SO(n)$.

Eqs.~\p{1}, \p{4} mean that in a superconformal gauge the right--moving sector
of the heterotic geometry is inert under supersymmetry transformations, and
eqs.~\p{2}, \p{5}--\p{7} imply that (0,n) superspace is to be superconformally
flat \cite{cons}, \cite{t}.

Under infinitesimal superdiffeomorphisms $\delta z^{\cal M}=\Xi^{\cal M}(z)$,
local $SO(1,1)\times SO(n)$ tangent space rotations $L_{\cal A}^{~\cal B}$ and
super--Weyl transformations $W(z)$ the inverse supervielbeins $E_{\cal A}^{\cal
M}$ are transformed as follows:
 \begin{equation}\label{8}
\delta E^{\cal M}_{\cal A}=-L^{~\cal B}_{\cal A} E^{\cal M}_{\cal
B}+\nabla_{\cal A}\Xi^{\cal M},
\end{equation}
 \begin{eqnarray}\label{9}
\delta_{W}E^{\cal M}_{++}&=&-2WE^{\cal M}_{++},\nn
\delta_{W}E^{\cal M}_{--}&=&-2WE^{\cal M}_{--}-iE^{\cal M}_{-A}\nabla_{-A}W,\nn
\delta_{W}E^{\cal M}_{-A}&=&-WE^{\cal M}_{-A}.
\end{eqnarray}

Reducing the transformations \p{8}, \p{9} to that of supergravity
transformations and partially solving the constraints one may, at least
locally, represent the covariant derivatives $\nabla_{\cal A}=E_{\cal A}^{\cal
M}\partial_{\cal M}$
(which act on the scalar superfields) in the following form \cite{}:
 \begin{equation}\label{10}
\nabla_{-A}=E{D}_{-A}\equiv E(\partial_{-A}-iV^\mu_{-A}\partial_\mu),
\end{equation}
 \begin{equation}\label{11}
\nabla_{--}=E^2D_{--}\equiv E^2V^\mu_{--}\partial_{\mu}=E^2{1\over
n}D_{-A}V^\mu_{-A}\partial_\mu,
\end{equation}
 \begin{equation}\label{12}
\nabla_{++}=E^{-2}(D_{++}+2iH^{(-4)}D_{--}-D_{-A}H^{(-4)}D_{-A}),
\end{equation}
where $E$ and $H^{(-4)}$ are independent superfields, $V^\mu_{-A}$ are subject
to constraints \cite{bstv}
 \begin{equation}\label{13}
D_{-A}V^\mu_{-B}+D_{-B}V^\mu_{-A}={2\over n}\delta_{AB}D_{-C}V^\mu_{-C},
\end{equation}
and derivatives $D_{-A},~D_{--},~D_{++}$ obey the global supersymmetry algebra
 \begin{equation}\label{14}
\{D_{-A},D_{-B}\}=-2iD_{--},\qquad [D_{-A},D_{\mp\mp}]=[D_{--},D_{++}]=0.
\end{equation}

For further constructing and studying the twistor string model it is essential
that in view of \p{8}--\p{13} one may choose a Wess--Zumino gauge in which
\begin{eqnarray}\label{15}
E&=&1,\qquad V^\mu_{-A}=\eta^{-}_Ae^\mu_{--}(\xi);\nn
D_{++}+2iH^{(-4)}D_{--}&=&e^\mu_{++}(\xi)\partial_{\mu}+
\eta^{-A}\Psi^{---}_A(\xi,\eta)e^\mu_{--}(\xi)\partial_{\mu},
\end{eqnarray}
where $e^\mu_{\pm\pm}(\xi)$ are $d=2$ graviton vielbeins, and the superfunction
$\Psi_{A}^{---}(\xi,\eta)$ (being a relic of $H^{(-4)}$) contains as components
a gravitino field $\psi_{A}^{---}(\xi)$, an
$SO(n)$ gauge field $A_{[AB]}^{--}(\xi)$ and fields $B_{[ABC...]}(\xi)$ which
may be identified as gauge fields of additional symmetries being part of $n>2$
local supersymmetry in $d=2$ \cite{12} (brackets imply antisymmetrization).

\section{$N=1$ twistor superstring action in (0,n) worldsheet superspace}

We use the derivatives \p{10}, \p{12} to construct the following doubly
twistorized $n=D-2$ superfield action of the $N=1$ superstring propagating (for
simplicity) in a flat target superspace parametrized by bosonic vector
coordinates $X^m$ and fermionic coordinates $\Theta^\alpha$ the latter being
Majorana or Majorana--Weyl spinors depending on the dimension D of space--time:
\begin{eqnarray}\label{16}
S&=&\int d^2\xi d^n\eta[({\rm
P}_{Am}(D_{-A}X^m-iD_{-A}\bar\Theta\gamma^m\Theta)\nn
&&+{\rm P}^{\cal MN}(i\partial_{\cal M}X^m\partial_{\cal
N}\bar\Theta\gamma_m\Theta-{1\over n}E^{++}_{\cal M}E^{--}_{\cal
N}D_{-A}\bar\Theta\gamma^mD_{-A}\Theta\bar\Lambda_{+}\gamma_m\Lambda_{+}+
\partial_{\cal M}Q_{\cal N})\nn
&&+{\rm P}_m(E^2\nabla_{++}X^m-iE^2\nabla_{++}\bar\Theta\gamma^m\Theta
-\bar\Lambda_{+}\gamma^m\Lambda_{+})],
\end{eqnarray}
where ${\rm P}_{mA},~{\rm P}^{\cal MN}=(-1)^{{\cal MN}+1}{\rm P}^{\cal NM}$,
and
${\rm P}_m$ are Lagrange multipliers, $Q_{\cal N}$ is an abelian gauge
worldsheet
superfield forcing the pullback into the worldsheet superspace of the
Wess--Zumino form $B=idX^m\wedge d\bar\Theta\gamma_m\Theta-{1\over
n}E^{++}\wedge E^{--}
D_{-A}\bar\Theta\gamma^mD_{-A}\Theta\bar\Lambda_{+}\gamma_m\Lambda_{+}$ to be a
closed form on the mass shell \cite{t}, \cite{dg}; and $\Lambda_+$ is a bosonic
spinor superfield whose space--time chirality may be the same or opposite to
that of $\Theta$. The action \p{16} is invariant under the following local
transformations
of $\Lambda_+$ (see, for example, \cite{sstr}):
\begin{equation}\label{16.}
\delta\Lambda^\alpha_+={1\over 2}(\bar\Lambda_{+}\gamma_m\Lambda_{+})(
\gamma_mU_-)^\alpha-(\bar\Lambda_+U_-)\Lambda^\alpha_+,
\end{equation}
where $U_-^\alpha$ is a spinor superfield parameter. Transformations \p{16.}
are
two--level reducible, so among $2D-4$ spinor components of $\Lambda_+$
only $D-1$ are independent, which is equal to the number of independent
components of the light--like vector.
Instead of $\Lambda_+$ we could alternatively use ``multitwistors''
$\Lambda_{+\dot A}$, where $\dot A$ are indices of an $n$--dimensional
representation of another local $SO(n)$ group. Then to reduce the number of
their components to that of the light--like vector one should include into the
action \p{16} the constraints \cite{n8p}, \cite{t}
\begin{equation}\label{16..}
\bar\Lambda_{+\dot A}\gamma^m\Lambda_{+\dot B}+\bar\Lambda_{+\dot
B}\gamma^m\Lambda_{+\dot A}={2\over n}\delta_{\dot A \dot B}\bar\Lambda_{+\dot
C}\gamma^m\Lambda_{+\dot C}.
\end{equation}

To analyze eq.~\p{16} it is convenient to rewrite it in a form being closer to
a twistor $N=1$ superstring action studied previously in \cite{t}, \cite{dg},
\cite{bstv}
\begin{eqnarray}\label{17}
S&=&\int d^2\xi d^n\eta[({\rm
P}_{Am}(D_{-A}X^m-iD_{-A}\bar\Theta\gamma^m\Theta)+{\rm P}^{\cal
MN}(\partial_{\cal M}Q_{\cal N}\nn
&&+i\partial_{\cal M}X^m\partial_{\cal N}\bar\Theta\gamma_m\Theta-{E^2\over
n}E^{++}_{\cal M}E^{--}_{\cal
N}D_{-A}\bar\Theta\gamma_mD_{-A}\Theta(\nabla_{++}X^m-
i\nabla_{++}\bar\Theta\gamma^m\Theta))\nn
&&+\hat{\rm P}_m(E^2\nabla_{++}X^m-iE^2\nabla_{++}\bar\Theta\gamma^m\Theta
-\bar\Lambda_{+}\gamma_m\Lambda_{+})],
\end{eqnarray}
where $\hat{\rm P}_m$ is a redefined Lagrange multiplier of \p{16}.

The first two lines in \p{17} constitute the action discussed in detail in
\cite{dg}. From the analysis performed  therein we know that the twistor
solution to the Virasoro constraint
$(\nabla_{--}X^m-i\nabla_{--}\bar\Theta\gamma^m\Theta)^2|_{\eta=0}=0$
corresponding to the right--moving sector of the $N=1$ superstring arises as a
consequence of the equation of motion of ${\rm P}_{mA}$ (which turns out to be
a completely auxiliary variable) and the constraints \p{6}:
\begin{eqnarray}\label{18}
&&D_{-A}X^m-iD_{-A}\bar\Theta\gamma^m\Theta=0,\nn
&&\nabla_{--}X^m-i\nabla_{--}\bar\Theta\gamma^m\Theta=
D_{-C}\bar\Theta\gamma^mD_{-C}\Theta,\\
&&D_{-A}\bar\Theta\gamma^mD_{-B}\Theta+D_{-B}\bar\Theta\gamma^mD_{-A}\Theta=
{2\over n}\delta_{AB}D_{-C}\bar\Theta\gamma^mD_{-C}\Theta.\nonumber
\end{eqnarray}

As for the Lagrange multiplier $P^{\cal MN},$ upon fixing a gauge of
corresponding local symmetries and solving for relevant equations of motion
it contains only one     non--zero component
\begin{equation}\label{18.}
p^{\mu\nu}\eta^n=T\varepsilon^{\mu\nu}\eta^n,
\end{equation}
where $T$ is identified with string tension \cite{ten}, \cite{dg}.

The last term in \p{16},  \p{17} is introduced for getting the twistor solution
to the second Virasoro constraint
$(\nabla_{++}X^m-i\nabla_{++}\bar\Theta\gamma^m\Theta)^2|_{\eta=0}=0$:
\begin{equation}\label{19}
(\nabla_{++}X^m-i\nabla_{++}\bar\Theta\gamma^m\Theta)=
E^{-2}\bar\Lambda_{+}\gamma^m\Lambda_{+}.
\end{equation}
For the induced metric $G_{\mu\nu}= (\partial_\mu
X^m-i\partial_\mu\bar\Theta\gamma^m\Theta)(\partial_\nu
X_m-i\partial_\nu\bar\Theta\gamma_m\Theta)$ on the string worldsheet to be
non--degenerate require that
\begin{equation}\label{20}
(\bar\Lambda_{+}\gamma_m\Lambda_{+})(D_{-A}\bar\Theta\gamma^mD_{-A}\Theta)\neq
0.
\end{equation}
Otherwise we would get a twistor--like formulation of a null superstring
\cite{bstv}.

With this in mind we will show that the all auxiliary variables containing in
the last term of \p{17}, including the gravitino and $SO(n)$ field, are either
zero or expressed in terms of the superstring coordinates
$x^m(\xi)=X^m|_{\eta=0}$ and $\theta^\alpha(\xi)=\Theta^\alpha|_{\eta=0}$ or
their derivatives, and thus do not lead to appearance of superfluous dynamical
degrees of freedom in the theory.

To get equations of  motion from action \p{17} we may substitute for
$\nabla_{++}$ its form in the Wess--Zumino gauge \p{15}. Then, upon redefining
${\rm P}_{Am}$ (in view of \p{18}) one may reduce the last term in \p{17}
to the following form
\begin{equation}\label{21}
S_{\Lambda}=\int D^2\xi d^n\eta\hat{\rm P}_m\left(e^\mu_{++}(\partial_\mu
X^m-i\partial_\mu  \bar\Theta\gamma^m\Theta)+{i\over
n}\eta^{-A}\Psi^{---}_{A}D_{-B}\bar\Theta\gamma^mD_{-B}\Theta-
\bar\Lambda_+\gamma^m\Lambda_+\right).
\end{equation}

Note that $\Psi^{---}_{A}$ and $\Lambda_+$ do not contribute to the other parts
of action \p{17}.
Thus, varying \p{21} with respect to $\Lambda_+$ and $\Psi^{---}_{A}$ we get
\begin{equation}\label{22}
(\hat{\rm P}_m\gamma^m\Lambda_+)^\alpha=0 \Longrightarrow \hat{\rm
P}_m=\Phi(\xi,\eta)\bar\Lambda_+\gamma_m
\Lambda_+,
\end{equation}
\begin{equation}\label{23}
(\eta^{-A}\hat{\rm P}_m)D_{-B}\bar\Theta\gamma^mD_{-B}\Theta=0,
\end{equation}
where $\Phi(\xi,\eta)$ is a superfield factor.

Taking into account the non--degeneracy condition \p{20}, from \p{22} and
\p{23} we see that the only non--zero component in $\Phi(\xi,\eta)$ is $\eta^n
a_{(-4)}(\xi)$, and hence $\hat{\rm P}^m$ is reduced to
\begin{equation}\label{24}
\hat{\rm P}^m=\eta^n\hat
p^m_{--}=\eta^na_{(-4)}(\xi)\bar\lambda_+\gamma^m\lambda_+,
\end{equation}
where $\lambda_+=\Lambda_+|_{\eta=0}$. This means that the term \p{21} does not
contribute to the equations of motion of $X^m$ and $\Theta$ obtained from
\p{17} starting from the next to the leading components in $\eta$--expansion of
$X^m$ and $\Theta$. This allows one to apply the reasoning of refs.~\cite{t},
\cite{dg} and show that the only nontrivial components of ${\rm P}^m_{A}$ are
\begin{equation}\label{25}
{\rm
P}^m_{A}=\eta^{-A_2}...\eta^{-A_n}\varepsilon_{A_2...A_nB}(\delta_{AB}p^m_{++}+
p^m_{AB++})
\end{equation}
(where $p^m_{AB++}$ is symmetric and traceless with respect to $A,B$),
 and that all higher components of $X^m$ and $\Theta$ are expressed in terms of
their leading components.

Now, varying \p{17} or \p{21} over $\hat{\rm P}_m$ we get
\begin{equation}\label{26}
\partial_{++} X^m-i\partial_{++}\bar\Theta\gamma^m\Theta+{i\over
n}\eta^{-A}\Psi^{---}_{A}D_{-B}\bar\Theta\gamma^mD_{-B}\Theta=
\bar\Lambda_+\gamma^m\Lambda_+\qquad (\partial_{\pm\pm}\equiv e^\mu_{\pm\pm}
\partial_\mu).
\end{equation}
This superfield equation can be solved by an iteration procedure.
At $\eta^{-A}=0$ we obtain the twistor solution to the second Virasoro
constraint:
\begin{equation}\label{27}
\partial_{++} x^m-i\partial_{++} \bar\theta\gamma^m\theta=\bar\lambda_+\gamma^m
\lambda_+.
\end{equation}
At the first order in $\eta$'s powers we have
\begin{equation}\label{28}
\partial_{++}\zeta^m_{-A}-i\partial_{++}\bar\theta\gamma^m\lambda_{-A}+
i\bar\theta\gamma^m\partial_{++}\lambda_{-A}+{1\over
n}\psi^{---}_{A}\bar\lambda_{-B}\gamma^m
\lambda_{-B}=2\bar\lambda_+\gamma^ms_{A},
\end{equation}
where $\lambda_{-A}=D_{-A}\Theta|_{\eta=0},~s_{A}=D_{-A}\Lambda_+|_{\eta=0}$,
and $\zeta^m_{-A}=D_{-A}X^m|_{\eta=0}=i\bar\theta\gamma^m\lambda_{-A}$, which
follows from equations of motion \p{18}.
Multiplying \p{28} by $\bar\lambda_+\gamma^m
\lambda_+$ and taking into account \p{20} we find that the gravitino field is
expressed through target superspace coordinates $~\theta$ and twistor--like
commuting spinors $\lambda_+,~\lambda_{-A}$:
\begin{equation}\label{29}
\psi^{---}_{A}=-2i{{(\partial_{++}\bar\theta\gamma^m\lambda_{-A})
(\bar\lambda_+\gamma_m\lambda_+)}\over{(\bar\lambda_+\gamma^n\lambda_+)
(\bar\lambda_{-B}
\gamma_n\lambda_{-B})}}.
\end{equation}

Substituting \p{29} back into eq.~\p{28} we have the sufficient number of
equations to express $s_A$ in terms of $x^m,~\theta,~\lambda_+$ and
$\lambda_{-A}$.
This is possible due to the local symmetry \p{16.} allowing one to eliminate
all extra degrees of freedom of $\Lambda_+$.

Proceeding further to the third  $\eta$'s order we can find the expression for
the $SO(n)$ gauge field $A^{--}_{[AB]}$ and $D_{-A}D_{-B}\Lambda_+|_{\eta=0}$
component of
$\Lambda_+$ in terms of the leading components of $X^M,~\Theta$ and
$\Lambda_+$. Following this way one may convince oneself that $\Psi^{---}_A $
and $\Lambda_+$ do not contain independent components. It means that adding the
new term to the superstring action \p{16} or \p{17} does not lead to extra,
undesirable, degrees of freedom and the model still remains equivalent to a
classical $N=1$ Green--Schwarz superstring.

Indeed, because all components of the superfield Lagrange multipliers ${\rm
P}^{\cal MN},~\hat{\rm P}_m$ and ${\rm P}_{mA}$ except \p{18.}, \p{24} and
\p{25} vanish, we may integrate \p{17}, or \p{16} over $\eta$ and get a
component action
$$
S=\int d^2\xi
e[p_{m--}(\partial_{++}x^m-i\partial_{++}\bar\theta\gamma^m\theta-
\bar\lambda_+\gamma^m\lambda_+)+p_{m++}(\partial_{--}x^m-
i\partial_{--}\bar\theta\gamma^m\theta-
{1\over n}\bar\lambda_{-A}\gamma^m\lambda_{-A})
$$
\begin{equation}\label{30}
+ {1\over
e}Ti\varepsilon^{\mu\nu}
\partial_{\mu}x^m\partial_{\nu}\bar\theta\gamma_m\theta+
{1\over
n}(\bar\lambda_{-A}\gamma^m\lambda_{-A})(\bar\lambda_+\gamma_m\lambda_+)
 + p^m_{AB++}\bar\lambda_{-A}\gamma_m\lambda_{-B}],
\end{equation}
where $e=det(e^{\pm\pm}_{\mu})$, and $p_{m\pm\pm},~p^m_{AB++}$ are
appropriately redefined Lagrange multipliers.

Varying \p{30} over $p_{m\pm\pm},~e^\mu_{\pm\pm}$ and $\lambda_{\pm}$ one finds
that
\begin{equation}\label{31}
p^m_{++}=\bar\lambda_+\gamma^m\lambda_+, \qquad p^m_{--}={1\over
n}\bar\lambda_{-A}\gamma^m\lambda_{-A}
\end{equation}
(from which it follows that $a_{(-4)}$ in eq.~\p{24} is zero).
Substituting \p{31} back into eq.~\p{30} we may get rid of $p^m_{\pm\pm}$ and
obtain a form of the twistor--like action for the $N=1$ superstring analogous
to that considered in \cite{ps}, \cite{bz}:
\begin{eqnarray}\label{31.}
S&=&\int d^2\xi e[{1\over
n}\bar\lambda_{-A}\gamma_m\lambda_{-A}(\partial_{++}x^m-
i\partial_{++}\bar\theta\gamma^m\theta)+
\bar\lambda_+\gamma_m\lambda_+(\partial_{--}x^m-
i\partial_{--}\bar\theta\gamma^m\theta)\nn
&& +{1\over e}
Ti\varepsilon^{\mu\nu}\partial_{\mu}x^m\partial_{\nu}\bar\theta\gamma_m\theta
-{1\over
n}(\bar\lambda_{-A}\gamma^m\lambda_{-A})(\bar\lambda_+\gamma_m\lambda_+)
 + p^m_{AB++}\bar\lambda_{-A}\gamma_m\lambda_{-B}].
\end{eqnarray}

This form of the string action turns out to be convenient for the
straightforward incorporation of heterotic fermions into the dynamics of the
model at the component level. Just add to \p{31.} a term:
\begin{equation}\label{32}
S_{\psi^I}=\int d^2\xi e \psi^I_{+}e^\mu_{--}\partial_\mu\psi^I_{+},
\end{equation}
where index $I$ corresponds to an internal symmetry group, which is $SO(32)$ or
$E_8\times E_8$ for the conventional $D=10$ heterotic string because of quantum
consistency reasons \cite{gsw}.

The contribution of the heterotic fermions \p{32} modifies the corresponding
left--moving Virasoro constraint
\begin{equation}\label{33}
(\partial_{++}x^m-i\partial_{++}\bar\theta\gamma^m\theta)^2+
\psi^I_{+}\partial_{++}\psi^I_{+}=0.
\end{equation}
A twistor--like solution to eq.~\p{33} which directly follows from action
\p{31.} plus \p{32} is
\begin{equation}
(\partial_{++}x^m-i\partial_{++}\bar\theta\gamma^m\theta)+{1\over
2}{{\bar\lambda_{-A}\gamma^m\lambda_{-A}}
\over{(\bar\lambda_{-B}\gamma^n\lambda_{-B})
(\bar\lambda_+\gamma_n\lambda_+)}}\psi^I_{+}\partial_{++}\psi^I_{+}
=\bar\lambda_+\gamma^m\lambda_+.
\end{equation}

\section{Conclusion}
 We have constructed the doubly twistorized version of the classical $N=1$
 superstring in $d=2, ~n=D-2$ worldsheet superspace which at least at the
component level can be generalized to include the chiral fermions. To write
down the action  we introduced in eq.~\p{16} the terms containing the bosonic
spinor superfield $\Lambda_+$. By somewhat ``mysterious'' reasons a bulk of
variables involved in the $n=D-2$ superfield description either vanish or are
expressed through the relevant dynamical characteristics of the string on the
mass shell. This happens, in particular, to the gravitino and other gauge
fields of $d=2,~n=D-2$ supergravity entering the action. So it is not required
to use extended supersymmetry transformations to gauge these fields away, and
the former may be used for matching the number of bosonic and fermionic
coordinates of the superstring itself, the role played by the fermionic
$\kappa$--symmetry in the conventional superstring approach.

As we have already mentioned, in \p{16} we could equally well use a set of
spinor superfields $\Lambda_{+\dot A}(\xi,\eta)$
($\dot A=1,...,D-2$) satisfying eq.~\p{16..}. In the latter case the model
considered may be regarded as a reduced version of a twistor--like $N=2$ closed
superstring (of type A or B), where $\Lambda_{+\dot A}$ is identified with
$D_{+\dot A}\Theta^2$, and $D_{+\dot A}$
is an odd covariant derivative acting in a right--moving sector of an enlarged
$(D-2,D-2)$ worldsheet superspace of the $N=2$ superstring (with $\Theta^2$
corresponding to the second space--time supersymmetry). For $D=3,$ type IIA
twistor superstring of \cite{gs2} this is indeed  the case. It seems of
interest to establish the analogous relationship between the present model and
that of ref.~\cite{ptm}.

\bigskip
{\bf Acknowledgments}. M.C. and D.S. are thankful to Lars Brink, Gabriele
Ferretti, Bengt E.W. Nilsson, Christian Preitschopf and Mario Tonin for
the interest
to this work and fruitful discussion. D.S. is grateful to the theoretical group
of Chalmers University of Technology for kind hospitality during his stay in
Gothenburg.

The work of I.B., D.S. and D.V. was partially supported by the American
Physical Society and International Science Foundation.


\newpage
\begin{thebibliography}{99}
\bibitem{ks}
J. A. De Azcarraga and J. Lukierski, {\sl Phys. Lett.} {\bf 113B} (1982) 170;\\
W. Siegel, {\sl Phys. Lett.} {\bf 128B} (1993) 397; {\sl Class. Quant. Grav.}
{\bf 2} (1985) 170.
\bibitem{bs}
R. Casalbuoni, {\sl Nuovo Cim.} {\bf 33A} (1976) 369;\\
A. I. Pashnev and D. V. Volkov, {\sl Teor. Mat. Fiz.} {\bf 44} (1980) 321;\\
L. Brink and J. H. Schwarz, {\sl Phys. Lett.} {\bf 100B} (1981) 310.
\bibitem{gsw}
M. B. Green, J. H. Schwarz and E. Witten, Superstring
Theory, CUP, 1987 (and references therein).
\bibitem{stv}
D. P. Sorokin, V. I. Tkach and
D. V. Volkov, {\sl Mod. Phys. Lett.} {\bf A4} (1989) 901.
\bibitem{lli}
R. Penrose {\sl J. Math. Phys.}
{\bf 8} (1967) 345; R. Penrose and M. A. H. MacCallum, {\sl Phys. Rep.} {\bf
6} (1972) 241, and refs. therein;\\
E. Witten, {\sl Nucl. Phys.} {\bf B226} (1986) 245;\\
L.-L. Chau and B. Milewski, {\sl Phys. Lett.} {\bf 216B} (1989) 330;\\
L.-L. Chau and C.-S. Lim, {\sl Int. J. Mod. Phys.} {\bf A4} (1989) 3819:\\
J. A. Shapiro and C. C. Taylor, {\sl Phys. Rep.} {\bf 191} (1990) 221.
\bibitem{stw}
A. Ferber, {\sl Nucl. Phys.} {\bf B132} (1978) 55;\\
T. Shirafuji, {\sl Progr. Theor. Phys.} {\bf 70} (1983) 18;\\
W. T. Shaw, {\sl Class. Quantum. Grav.} {\bf 3} (1986) 753;\\
P. Budinich, {\sl Comm. Math. Phys.} {\bf 107} (1986) 455;\\
A. K. H. Bengtsson, I. Bengtsson, M. Cederwall and E. Linden, {\sl Phys. Rev.}
{\bf 36D} (1987) 1786;\\
I. Bengtsson and M. Cederwall, {\sl Nucl. Phys.} {\bf B302} (1988) 81;\\
Y. Eisenberg and S. Solomon, {\sl Nucl. Phys.} {\bf B309} (1988) 709;\\
M. Cederwall, {\sl Phys. Lett.} {\bf 226B} (1989) 45;\\
M. Plyushchay, {\sl Mod. Phys. Lett.} {\bf A4} (1989) 1827;\\
D. P. Sorokin, {\sl Fortsch. der Phys.} {\bf 38} (1990) 302;\\
A. I. Gumenchuk and D. P. Sorokin, {\sl Sov. J. Nucl. Phys.} {\bf
51} (1990) 350.
\bibitem{harm}
E. Sokatchev, {\sl Class. Quantum Grav.} {\bf 4} (1988) 237;\\
E. Nissimov, S. Pacheva and S. Solomon, {\sl Nucl. Phys.} {\bf B296} (1988)
469; {\bf B297} (1988) 349; {\bf B299} (1988) 183; {\bf B317} (1988) 344;
{\sl Int. J. Mod. Phys.} {\bf A4} (1989) 737; \\
R. E. Kallosh and M. A. Rahmanov, {\sl Phys. Lett.} {\bf 209B} (1988) 233;
{\bf 214B} (1988) 549;\\
A. S. Galperin, P. S. Howe and K. S. Stelle, {\sl Nucl. Phys.} {\bf
B368} (1992) 281;\\
F. Delduc, A. Galperin and E. Sokatchev, Ibid.
{\bf B368} (1992) 143.
\bibitem{bz}
I. A. Bandos, {\sl Sov. J. Nucl. Phys.} {\bf 51} (1990) 906;\\
I. A. Bandos and A. A. Zheltukhin,
{\sl JETP Lett.} {\bf 51} (1990) 547; {\bf 54} (1991) 421;
{\bf 55} (1992) 81;
{\sl Phys. Lett.} {\bf B261} (1991) 245; {\bf B288} (1992) 77;
{\sl Theor. Math. Phys.} {\bf 88} (1991) {358};
{\sl Fortsch. der Phys.} {\bf 41} (1993) 619;
 {\sl Sov. J. Nucl. Phys.} {\bf 56} (1993) 198;
{\sl Int. J. Mod. Phys.} {\bf A8} (1993) 1081.
\bibitem{spa}
D. V. Volkov and A. A. Zheltukhin, {\sl Lett. Math. Phys} {\bf 17} {(1989})
{141};\\
D. P. Sorokin, V. I. Tkach, D. V. Volkov and A. A. Zheltukhin,
{\sl Phys. Lett.} {\bf 216B} (1989) 302;\\
 F. Delduc and E. Sokatchev, {\sl Class. Quantum  Grav.} {\bf
9} (1991) 361;\\
E. A. Ivanov
and A. A. Kapustnikov, {\sl Phys. Lett.} {\bf 267B} (1991) 175;\\
J. P. Gauntlett, {\sl Phys. Lett.} {\bf B272} {(1991)} {25};\\
M. Cederwall, A note on the relation between different forms of superparticle
dynamics. Preprint Gothenburg-ITP-93-33, 1993, hep-th/9310177.
\bibitem{ps}
V. A. Soroka, D. P. Sorokin, V. I. Tkach and D. V. Volkov,
{\sl Int. J. Mod. Phys.} {\bf A7} (1992) 5977;\\
A. I. Pashnev and D. P. Sorokin, {\sl Class. Quantum Grav.} {\bf 10} {(1993)}
{625}.
\bibitem{n8p}
A. Galperin and E. Sokatchev, {\sl Phys. Rev.} {\bf
D46} (1992) 714.
\bibitem{bstv}
I. Bandos, D. Sorokin, M. Tonin and D. Volkov,
{\sl Phys. Lett.} {\bf 319B} (1993) 445.
\bibitem{sstr}
N. Berkovits, {\sl Phys. Lett.} {\bf
232B} (1989) 184; {\bf 241B} (1990) 497; {\sl Nucl. Phys.} {\bf B350}
(1991) 193; {\bf B358} (1991) 169; {\sl Nucl. Phys.} {\bf B379} (1992) 96; {\bf
B395} (1993) 77.
\bibitem{t}
M. Tonin, {\sl Phys. Lett.} {\bf 266B}
(1991) 312; {\sl Int. J. Mod. Phys.} {\bf A7} (1992) 6013;\\
S. Aoyama, P.
Pasti and M. Tonin, {\sl Phys. Lett.} {\bf 283B} (1992) 213;\\
\bibitem{dis}
F. Delduc,
E. Ivanov and E. Sokatchev, {\sl Nucl. Phys.} {\bf B384} (1992) 334;\\
\bibitem{dg} F. Delduc, A. Galperin, P. Howe and E. Sokatchev, {\sl Phys.
Rev.} {\bf D47} (1992) 578.
\bibitem{st}
D. P. Sorokin and M. Tonin, On the chiral fermions in the twistor--like
formulation of D=10 heterotic string. Preprint DFPD/93/TH/52, Padova, 1993.
Hep--th/9307195. Phys. Lett. B (in press).
\bibitem{vz}
D. V. Volkov and A. A. Zheltukhin,  {\sl Nucl. Phys.} {\bf B335} (1990) 723.
\bibitem{cp}
V. Chikalov and A. Pashnev, {\sl Mod. Phys. Lett.} {\bf A8} (1993) 285.
\bibitem{ptm}
P. Pasti and M. Tonin, Twistor--like formulation of the supermembrane in
$D=11$. Preprint DFPD/93/TH/07, Padova,  1993. Nucl. Phys. B (in press).\\
Twistor--like formulation of $D=10$, type IIA superstrings. \\Preprint
DFPD/94/TH/05, Padova, 1994.
\bibitem{gs2}
A. Galperin and E. Sokatchev, {\sl Phys. Rev.} {\bf D48} (1993) 4810.
\bibitem{cons}
P. S. Howe, {\sl J. Phys.} {\bf A12} (1979) 393;\\
P. Nelson and G. Moore, {\sl Nucl. Phys.} {\bf B274} (1989) 509;\\
P. S. Howe and G. Papadopoulos, {\sl Class. Quantum Grav.} {\bf 4} (1987) 11.
\bibitem{12}
M. Ademollo, L. Brink, A. D'Adda et. al., {\sl Phys. Lett.} {\bf 62B} (1976)
105; {\sl Nucl. Phys.} {\bf B114} (1976) 297.
\bibitem{ten} J. A. De Azc\`arraga, J. M. Izquierdo and P. K.
Townsend, {\sl Phys. Rev.} {\bf 45} {(1992)} {3321};\\ P. K. Townsend, {\sl
Phys. Lett.} {\bf B277} ({1992}) {285};\\
E. Bergshoeff, L. A. J. London and P. K.Townsend,  {\sl Class. Quantum Grav.}
{\bf 9} ({1992}) {2545}.
\end{thebibliography}
\end{document}